\documentclass[superscriptaddress,pre,showpacs,preprintnumbers,amsmath,amssymb,aps]{revtex4}
\usepackage{natbib}
\usepackage{graphicx}% Include figure files
\usepackage{dcolumn}% Align table columns on decimal point
\usepackage{bm}% bold math
\usepackage{color}
\usepackage{subfig}

\begin{document}

%\preprint{Submitted to Physical Review Letters, Version 2}
%\preprint{Submitted to Physical Review Letters}
%\preprint{Submitted to \red{PRE ?}...}

\title{Comment on the paper\\
J. G. Zhou, Rectangular lattice Boltzmann method, \\Phys. Rev. E 81, 026705 (2010)}% Force breaks \\

\author{Shyam Chikatamarla}
\affiliation{Aerothermochemistry and Combustion Systems Lab, ETH Zurich, 8092 Zurich, Switzerland}

\author{Ilya Karlin}
\affiliation{Aerothermochemistry and Combustion Systems Lab, ETH Zurich, 8092 Zurich, Switzerland}
\affiliation {School of Engineering Sciences, University of Southampton, SO17 1BJ Southampton, UK}

\date{\today}% It is always \today, today,

\begin{abstract}
{It is shown both analytically and numerically that the suggested lattice Boltzmann model on rectangular grids leads to anisotropic dissipation of fluid momentum and thus it does not recover Navier-Stokes equations. Hence, it cannot be used for the simulation of hydrodynamics.}
\end{abstract}


\maketitle

In a recent paper \cite{Zhou2010}, a lattice Bhatnagar-Gross-Krook (LBGK) model is suggested for the simulation of the Navier-Stokes equation for incompressible flow. Setting itself apart from the
standard LBGK equation \cite{Succi} on square lattices, the LBGK model of Ref.\ \cite{Zhou2010} is formulated on generic rectangular lattices. This is surprising since earlier studies clearly indicated that it is impossible to formulate a single relaxation time LBGK model for the Navier-Stokes equation on simple rectangular lattices \cite{Bouzidi2001}.

However, a more careful study of the LBGK model in \cite{Zhou2010} detailed below reveals that the resulting hydrodynamic equations, in fact, lead to anisotropic
viscous pressure tensor, thus disproving its main result. Indeed, following \cite{Zhou2010}, let us consider a rectangular two-dimensional lattice with the velocities $\bm{e}_{\alpha}=(e_{\alpha x},e_{\alpha y})$, $\alpha=0,\dots,8$, where $e_{\alpha x}\in\{-e_x,0,e_x\}$ and
$e_{\alpha y}\in\{-e_y,0,e_y\}$ (the rectangular D2Q9 lattice, see Fig.\ 1 in \cite{Zhou2010}), and
revisit the derivation of the Navier-Stokes equation from the LBGK model presented in Appendix B of \cite{Zhou2010}. Specifically, the computation of the non-equilibrium pressure tensor $\bm{\Pi}^{(1)}$ (B23), (B24) amounts to evaluation of a function which appears in the left hand side of equation (B21),
\begin{equation}
\frac{\partial}{\partial x_k}\sum_{\alpha}e_{\alpha i}e_{\alpha j}e_{\alpha k}f^{(0)}_{\alpha}.
\end{equation}
Let us denote $Q^{\rm (0)}_{ijk}$ the third-order moment at equilibrium in the above expression,
\begin{equation}\label{eq:third}
Q^{\rm (0)}_{ijk}=\sum_{\alpha}e_{\alpha i}e_{\alpha j}e_{\alpha k}f^{(0)}_{\alpha},
\end{equation}
where $f^{(0)}$ is the equilibrium population $f^{\rm eq}$ given by equation (11) of \cite{Zhou2010}. For the two-dimensional model there are four independent components of the tensor (\ref{eq:third}), $Q^{(0)}_{xxx}$, $Q^{(0)}_{yyy}$, $Q^{(0)}_{xyy}$ and $Q^{(0)}_{yxx}$. The two diagonal components, $Q^{(0)}_{xxx}$ and $Q^{(0)}_{yyy}$, are independent of any particular form of the equilibrium because of the lattice constraint, $e_{\alpha i}^3=e_i^2e_{\alpha i}$. Therefore, for the diagonal components we have
$Q_{xxx} = \rho e_x^2 u_x$ and $Q_{yyy} = \rho e_y^2 u_y$, irrespectively of the particular form of $f^{\rm eq}$, or, introducing the aspect ratio  $a=e_x/e_y$,
\begin{align}
\begin{split}
\label{eq:moments1}
Q^{(0)}_{xxx} = \rho e_x e_y a u_x,\
Q^{(0)}_{yyy} = \rho e_x e_y a^{-1}u_y
\end{split}
\end{align}
On the other hand, evaluating the remaining two off-diagonal components on the equilibrium (11), we find
\begin{align}
\begin{split}
\label{eq:moments2}
Q^{(0)}_{xyy} =
\frac{\rho e_x e_y}{3} a^{-1}u_x,\
Q^{(0)}_{yxx} =
\frac{\rho e_x e_y}{3} au_y.
\end{split}
\end{align}
{Note that equations (\ref{eq:moments1}) and (\ref{eq:moments2}) are \emph{exact}.}
Thus, unless $a=1$, tensor $Q^{\rm (0)}_{ijk}$ (\ref{eq:third}) is not isotropic, and hence the right hand side of (B21) as given in \cite{Zhou2010} is not correct. When the correct expressions (\ref{eq:moments1}) and (\ref{eq:moments2}) are used in the further steps of the derivation, we finally obtain, at variance with the isotropic Newtonian viscous stresses (B24), an anisotropic expression depending on the aspect ratio $a$,
\begin{widetext}
\begin{equation}\label{eq:stress}
\bm{\Pi}^{(1)}=\nu\left[
 \begin{array}{cc}
 a\left(3\frac{\partial (\rho u_x)}{\partial x}+\frac{\partial (\rho u_y)}{\partial y}\right) &
 a\frac{\partial (\rho u_y)}{\partial x}+\frac{1}{a}\frac{\partial (\rho u_x)}{\partial y}\\
 a\frac{\partial (\rho u_y)}{\partial x}+\frac{1}{a}\frac{\partial (\rho u_x)}{\partial y} &
 \frac{1}{a}\left(3\frac{\partial (\rho u_y)}{\partial y}+\frac{\partial (\rho u_x)}{\partial x}\right)\\
                       \end{array}
                     \right].
                     \end{equation}
\end{widetext}
where $\nu$ is given by equation (B25) in Ref.\ \cite{Zhou2010}.
If $a=1$ (standard square grid), the viscous stresses (\ref{eq:stress}) become isotropic, and $\nu$ is interpreted as the kinematic viscosity. However, there is no such interpretation at $a\ne1$, and (\ref{eq:stress}) is not an isotropic viscous stress of the Navier-Stokes equation.
{Thus, the standard analysis of hydrodynamic limit reveals that the model \cite{Zhou2010} \emph{does not} recover Navier-Stokes equations at $a\ne1$.}

Same consideration applies to the three-dimensional case of \cite{Zhou2010}. It is easy to see that a particular choice of the equilibrium is actually irrelevant because the anisotropy is already present in the diagonal components of the third-order moment (\ref{eq:moments1}) which are independent of a particular choice of the equilibrium.

{To numerically probe the anisotropy in the model \cite{Zhou2010}, we study below simulations of the standard Taylor-Green (TG) vortex flow in two dimensions at Reynolds number ${\rm Re}=10$.
The simulation with a standard square lattice is fully resolved and agrees well with the analytical solution on the grid with $32\times32$ nodes.
However, when we refine the grid and increase the $x$ resolution to $64$ points while retaining the y resolution at $32$ points {(that is, $e_y=2e_x$)}, we clearly see the anisotropic behavior of viscosity.
The velocity in the $y$ direction decays much faster resulting in a deformed vortex which eventually corrupts the entire simulation. Fig.\ \ref{fig1} also shows that the anisotropy may go unnoticed in the early stages of the simulation (see for example the snapshot at $T=500$ in Fig.\ \ref{fig1}).
{
In the simulation, Reynolds number was defined as ${\rm Re}=U_0 N_x/\nu$, where $U_0$ is the amplitude of the velocity of TG vortex, $N_x$ is the number of nodes in the $x$ direction and $\nu$ is the parameter in (\ref{eq:stress}) related to the LBGK relaxation time $\tau$ according to equation (B25) of  \cite{Zhou2010}: $\nu=e_xe_y(2\tau-1)/6$.
}

{We note in passing that at $a=1$  the equilibrium (11) in \cite{Zhou2010} is different from the standard equilibrium used in LBGK models, in particular, equilibrium populations of the diagonal velocities become equal to zero at vanishing flow velocity. This probably explains why model \cite{Zhou2010} is numerically less stable than the standard LBGK model on a square lattice.
The simulation of Taylor-Green vortex flow on a grid of $32\times32$ remains stable for up to ${\rm Re}=10000$ for the standard lattice Boltzmann scheme while it fails below ${\rm Re}=150$ with the scheme \cite{Zhou2010}.}

\begin{figure}[htbp!]
%\centering
  \begin{tabular}{cc}
	   \subfloat[][$T=0$]{\includegraphics[width=0.45\textwidth]{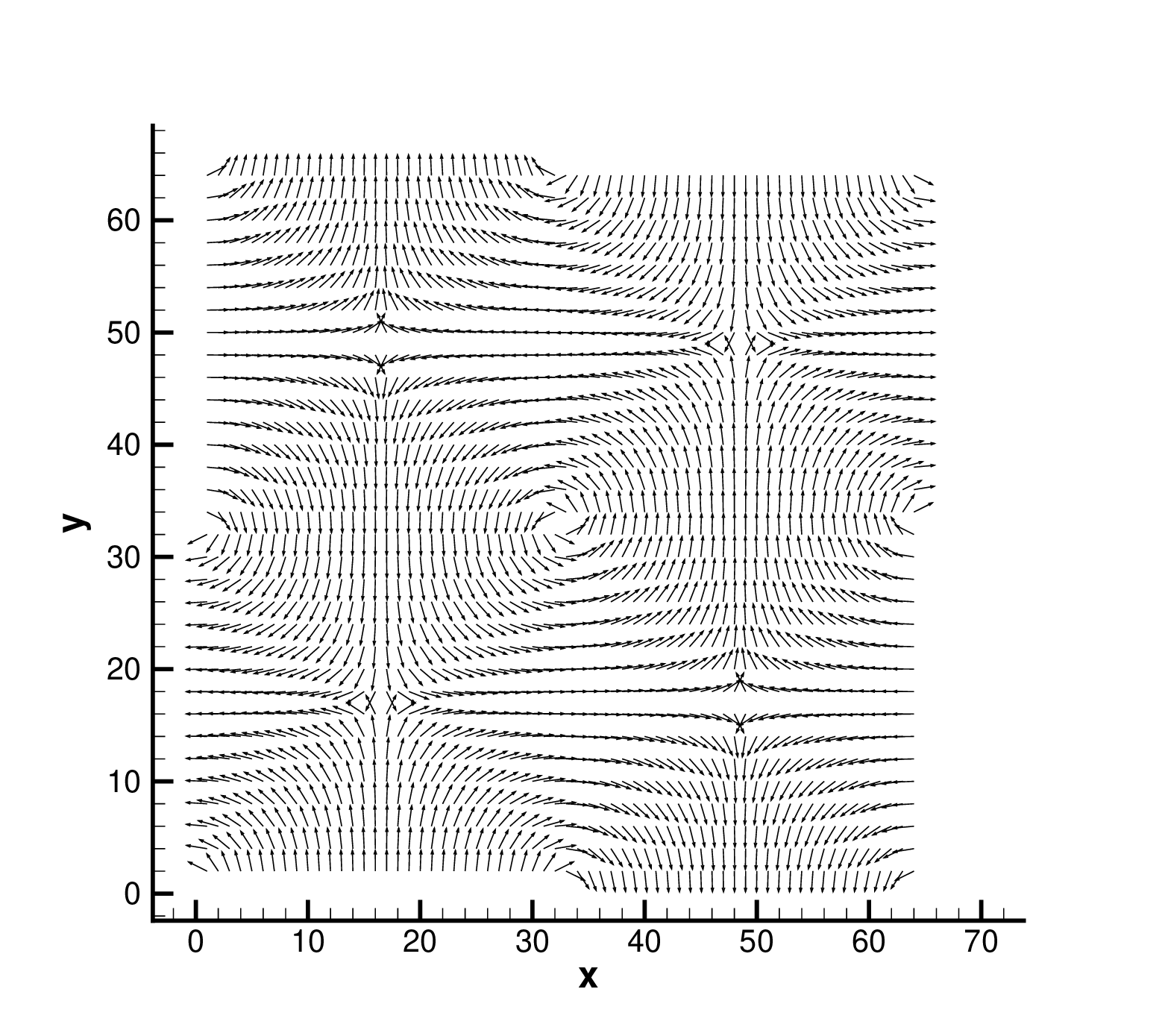} } &
        \subfloat[][$T=500$ ]{\includegraphics[width=0.45\textwidth]{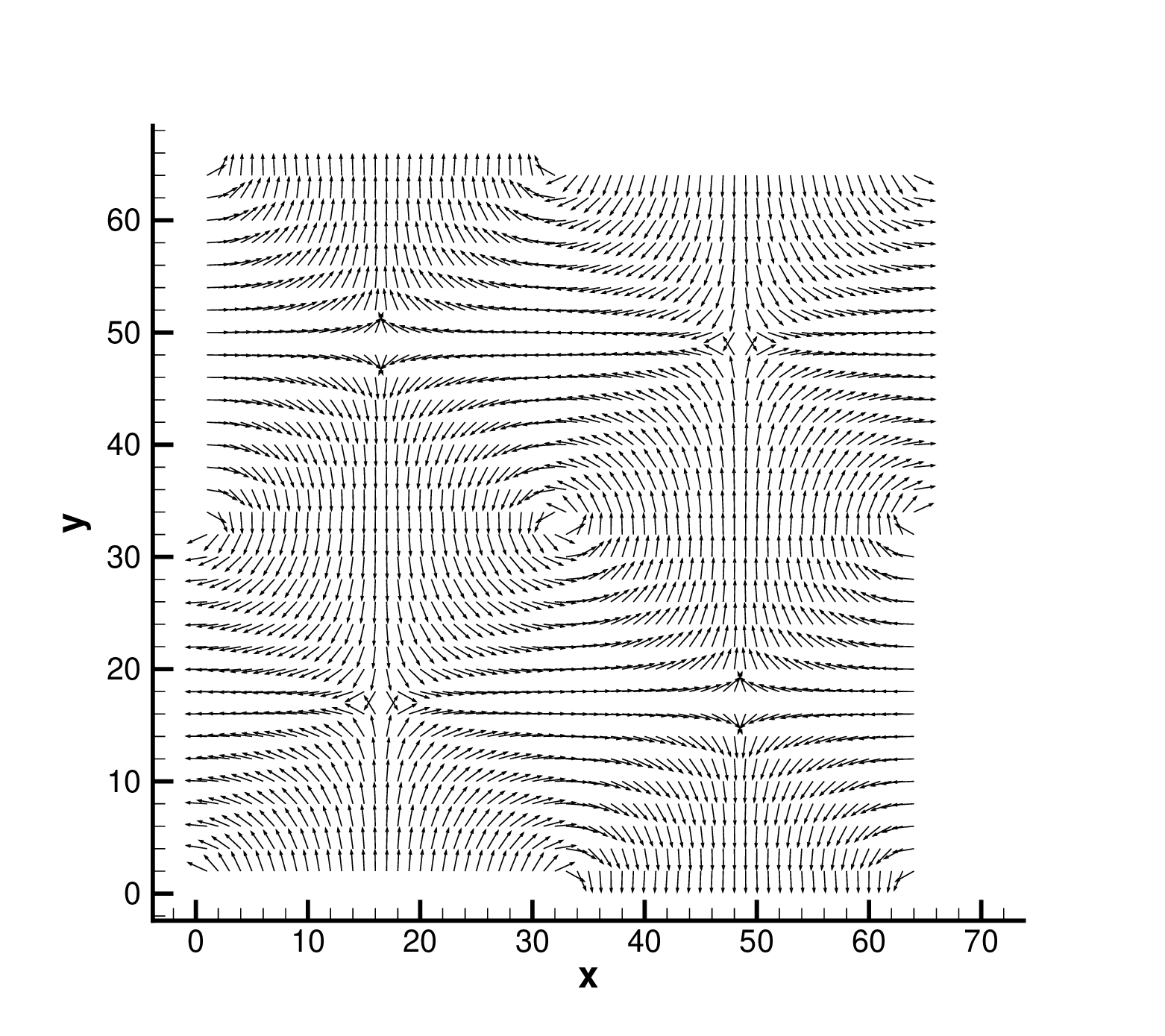} } \\
      \subfloat[][$T=1100$ ]{ \includegraphics[width=0.45\textwidth]{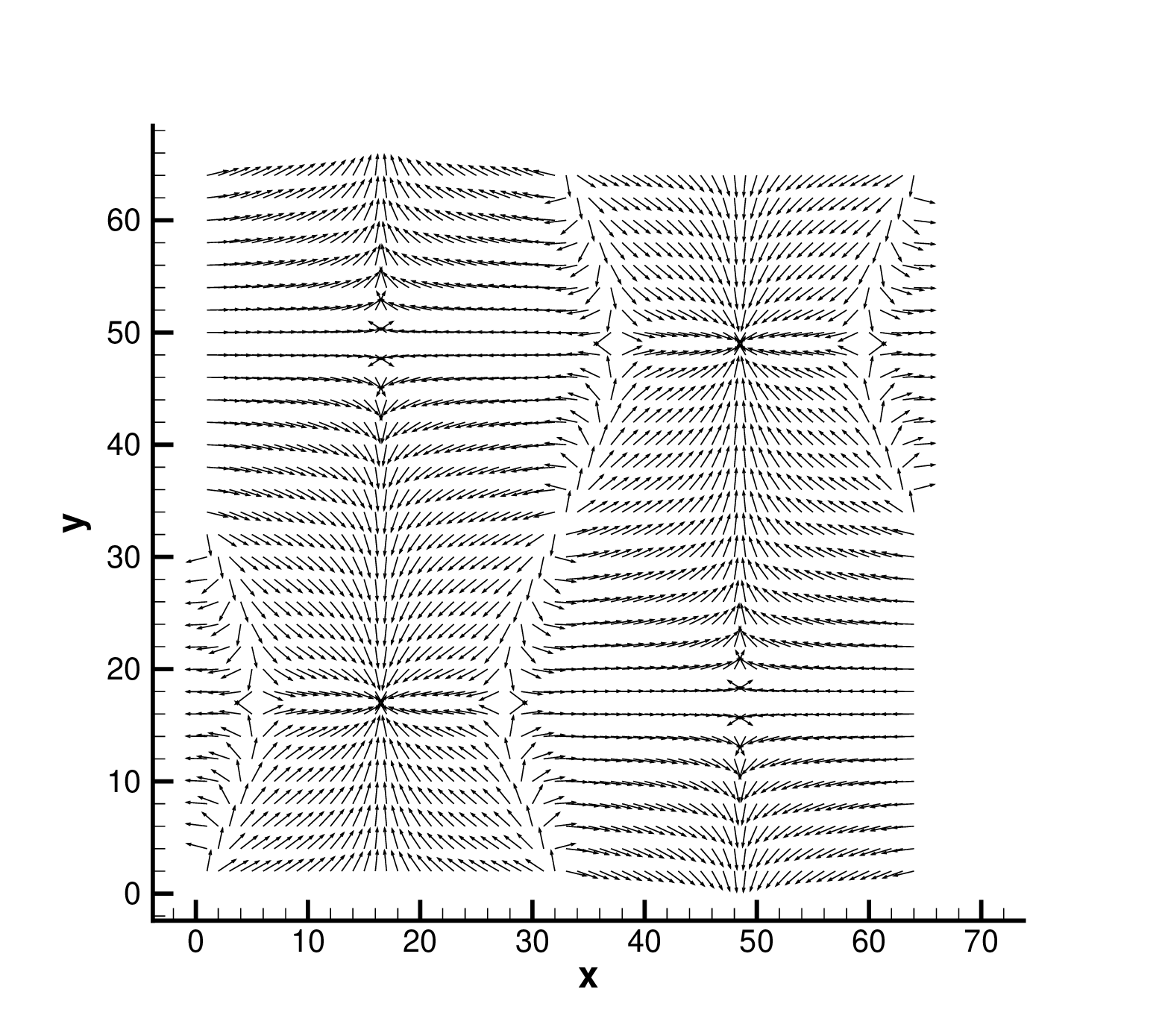}} &
       \subfloat[][ $T=1500$]{\includegraphics[width=0.45\textwidth]{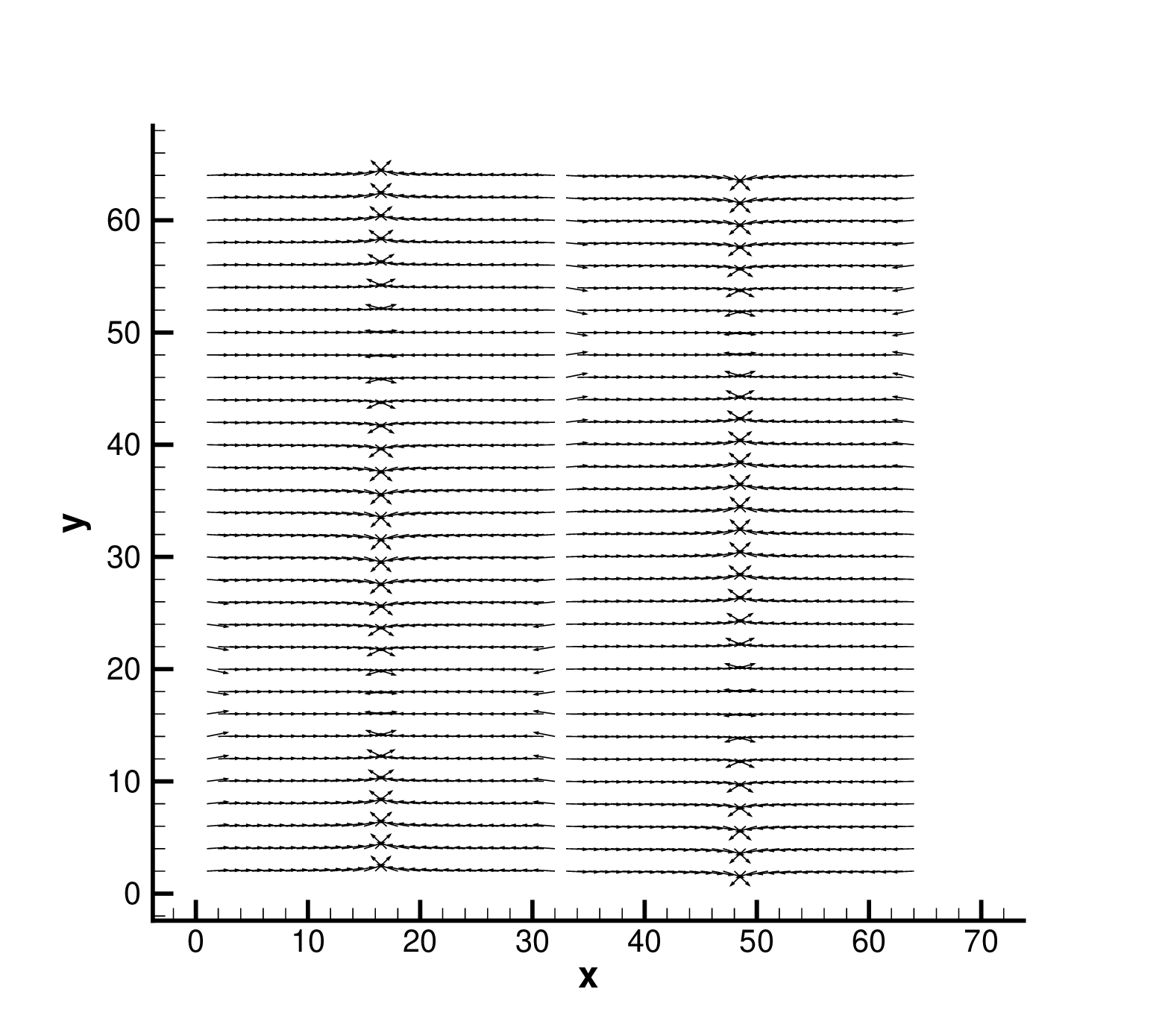}}
	\end{tabular}
\caption{Taylor-Green vortex simulation with model \cite{Zhou2010} at Reynolds number ${\rm Re}=10$, initial velocity $U_0=0.05$ and a grid of $N_x=64$, $N_y=32$ ($e_y=2e_x$), $\nu=U_0\times N_x/{\rm Re}=0.32$. Velocity vector plots (with uniform vector length) at various times show the accumulation of anisotropy: The vortex is visibly corrupted at $T=1100$ and is fully destroyed at $T=1500$.}
\label{fig1}
\end{figure}

%%%%%%%%%%%%%%%%%%%%%%%
{Finally, a comment on the simulations presented in \cite{Zhou2010} is in order. For some flow configurations (unidirectional flows) the above anisotropy may be not very visible in the velocity plots.
\emph{All} simulations with rectangular lattices in \cite{Zhou2010} concern predominantly unidirectional flows without vortex structures. Such flows do not provide a benchmark for an anisotropic model of \cite{Zhou2010} as they exclude rotation of the flow. Even the simplest vortex simulation reveals unphysical anisotropy of the model, as was shown above.}

{To conclude, the model suggested in \cite{Zhou2010} does not recover the Navier-Stokes equations for non-square rectangular lattices and  cannot be used for simulation of hydrodynamics.}

\end{document}